\documentclass[12pt]{article}
\usepackage{amssymb}

\begin{document}
\begin{titlepage}




{\hbox to\hsize{\hfill November 2011 }}
{\hbox to \hsize{\hfill Revised November 2012}}

\bigskip \vspace{3\baselineskip}

\begin{center}

{\bf \Large 

Electroweak scale invariant models with }
\vskip 0.4cm

{\bf \Large
small cosmological constant}

\bigskip

\bigskip

\bigskip

\bigskip

{\bf Robert Foot$^{\rm a}$ and Archil Kobakhidze$^{\rm a,b}$  \\}

\smallskip

\bigskip

{ \small \it

ARC Center of Excellence for Particle Physics at the Terascale, \\
$^{\rm a}$School of Physics, The University of Melbourne, Victoria 3010, Australia \\
$^{\rm b}$School of Physics, The University of Sydney, NSW 2006, Australia \\
E-mails: rfoot@unimelb.edu.au, archilk@unimelb.edu.au
\\}

\bigskip

\bigskip

\bigskip

\bigskip

\bigskip

{\large \bf Abstract}

\end{center}

\noindent
We consider scale invariant models where the classical scale invariance is broken perturbatively by radiative
corrections at the electroweak scale.   These models offer an elegant and simple solution to the hierarchy problem.
If we further require the cosmological constant to be small then such models are also highly predictive.
Indeed, the minimal such model, comprising a Higgs doublet and a real singlet, has the same
number of parameters as the standard model. Although this minimal model is disfavoured by recent
LHC data, we show that two specific extensions incorporating neutrino masses and dark matter 
are fully realistic. That is, consistent with all experiments and observations.
These models predict a light pseudo-Goldstone boson, $h$, with mass around 10 GeV or less. 
A fermionic-bosonic mass relation is also predicted.
The specific models considered, as well as more generic scale invariant models, can be probed at the LHC.

\end{titlepage}

\section{Introduction}

The quantum stability of the electroweak scale (hierarchy problem)  is an interesting motivation for new physics 
as it suggests enlarged symmetry of the Standard Model. In this context, 
scale invariance is a promising candidate for such an additional symmetry of particle interactions.  
Scale invariance, can be an exact classical symmetry,  
broken radiatively as a result of the quantum anomaly. 
This generates a mass scale through the quantum-mechanical phenomenon of dimensional transmutation\cite{Coleman:1973jx}. 
Despite its anomalous nature, scale invariance  still ensures the stability of the electroweak scale since 
the radiative breaking is ``soft" as it generates only logarithmic corrections to the electroweak scale\cite{bardeen,Coleman:1973jx}.  
Realistic scale invariant models broken perturbatively by quantum corrections can be constructed which typically 
feature a scalar sector consisting of the usual Higgs doublet 
together with a Higgs singlet and possibly other scalar fields\cite{Foot:2007as, Foot:2007ay, Foot:2007iy, Foot:2010av}. 
For related works see also\cite{Hempfling:1996ht, Chang:2007ki, Hambye:2007vf, 
Iso:2009ss, Holthausen:2009uc, AlexanderNunneley:2010nw, Latosinski:2010qm}.

It is possible that all scales: electroweak, neutrino mass, cosmological constant and Planck scales
originate radiatively via quantum corrections. In this article, we consider a class of 
scale invariant theories which we call `electroweak scale invariant models' whereby the `low energy'
effective Lagrangian describing physics below the Planck scale is scale invariant. 
The relevant particle physics scales such as the electroweak scale and the neutrino mass 
scale will arise through the dimensional transmutation at energies $\sim$ TeV. 

Quantum corrections in scale invariant models also generate a finite, and thus in principle calculable,   cosmological constant  (CC). 
Constraining the cosmological constant to be small
imposes interesting constraints on scale invariant models \cite{Foot:2010et}.
We first set aside the issue of neutrino mass and reconsider the simplest scale invariant electroweak model with a scalar content
consisting of a Higgs doublet and a real singlet\cite{Foot:2007as}. Scale invariance together with the small CC leads 
to a highly constrained theory with the same number of parameters as the standard model.
This model, although disfavoured by recent LHC data, is an important `toy model'.
We discuss two realistic extensions of this model. The first incorporates neutrino masses
via the type-II see-saw mechanism. The second one introduces a hidden sector to accommodate dark matter,
which we argue is a particularly natural framework in scale invariant models.

\section{Perturbatively small cosmological constant in scale invariant theories}

The incorporation of a small cosmological constant within scale invariant theories has
been discussed in Ref. \cite{Foot:2010et}. It will lead to important constraints on
realistic scale invariant theories, so we briefly review this material here.

Consider a classically scale-invariant theory with 
$n$ real scalar fields $S_i$ ($i=1,2,...n$).  
The classical potential can be expressed as:
\begin{equation}
V_0(S_i)=\lambda_{ijkl}S_iS_jS_kS_l~,
\label{a1}
\end{equation}
where summation over repeated indices is assumed and $\lambda_{ijkl}$ are bare coupling constants. 
It is useful to adopt the hyper-spherical parameterization for the scalar fields:
\begin{eqnarray}
S_i(x) &=& r(x)\cos\theta_i(x)\prod_{k=1}^{i-1}\sin\theta_{k}(x)~, \ {\rm for} \ i=1,\ldots,n-1 \nonumber \\
S_n (x) &=& r(x) \prod_{k=1}^{n-1} \sin\theta_k
\label{a2}
\end{eqnarray}
where $r(x)$ is the modulus field. If this field acquires
a nonzero vacuum expectation value (VEV), $\langle r \rangle \neq 0$, 
then scale invariance is spontaneously broken. The resulting (pseudo-)Goldstone boson  is the dilaton\cite{GildenerWeinberg,Coleman:1973jx}. 
With the above parameterization  the classical potential takes the form
\begin{equation}
V_0(r, \theta_i)=r^4f(\lambda_{ijkl}, \theta_i)~.
\label{a3}
\end{equation} 
The modulus field $r(x)$ factors out due to the classical scale invariance.
The extremum  condition 
$\left.\frac{\partial V_0}{\partial r}\right |_{r=\langle r \rangle,~ \theta_i=\langle \theta_i \rangle}=0$ 
implies that the VEV of the potential, that is, the classical contribution to the CC, vanishes: 
$V_0(\langle r\rangle, \langle \theta_i\rangle )=0$. At the classical level, the dilaton field 
remains massless and the VEV $\langle r \rangle=0$, unless $f(\lambda_{ijkl}, \theta_i)=0$ in which case  $\langle r \rangle$ 
is undetermined (flat direction).   

Including quantum corrections results in an effective potential which can be written in terms of couplings
and fields which depend on the   
renormalization scale $\mu$,
\begin{eqnarray}
V=A(g_a(\mu), m_{x}(\mu), \theta_i(\mu), \mu) r^{4}(\mu)+B(g_a(\mu), m_{x}(\mu), 
\theta_i(\mu), \mu) r^{4}(\mu)\log\left(\frac{r^2(\mu)}{\mu^2}\right) \nonumber \\
+C(g_a(\mu), m_{x}(\mu), \theta_i(\mu), \mu) r^{4}(\mu)\left[\log\left(\frac{r^2(\mu)}{\mu^2}\right)\right]^2+\ldots~,
\label{a4}
\end{eqnarray}
where $\ldots$ denotes all terms with  higher-power logarithms and
$g_a(\mu)$ and $m_x(\mu)$ denote all relevant 
running dimensionless couplings and effective masses. It is 
very convenient to set the renormalization scale as $\mu = \langle r \rangle$ because for this special choice of
$\mu$ the higher-power $\log$ terms become irrelevant for our discussion. 
We consider a fixed radial direction by taking 
$\theta_i = \langle \theta_i \rangle$ in  Eq.(\ref{a4}). The extremum condition along this radial direction is
\begin{eqnarray}
\frac{\partial V}{\partial r} =0 & \Rightarrow & 2A(\mu=\langle r\rangle)  + B(\mu=\langle r\rangle)  = 0 \ .
\label{1}
\end{eqnarray}
If we demand that the perturbative contribution to the CC vanishes\footnote{
In practice, to achieve a tiny CC within observational limits, the perturbative contribution to the CC should not vanish but approximately cancel the
small QCD contribution. However, having the CC at the QCD scale, rather than zero does not lead to any important modifications of our analysis.
See ref.\cite{Foot:2010et} for further discussions.}, 
then this requires 
\begin{eqnarray}
V_{\rm min} =0 & \Rightarrow & A(\mu=\langle r\rangle)  = 0~.
\label{2}
\end{eqnarray}
Note that while $V_{\min}=0$ implies tuning of parameters,
the condition Eq.(\ref{1}) simply defines 
$\langle r \rangle$ as the scale $\mu$
where $2A +  B = 0$. Hence, the condition Eq.(\ref{1}) effectively replaces one dimensionless 
parameter for a dimensional parameter, the phenomenon known as dimensional transmutation. 

The two conditions Eq.(\ref{1}) and Eq.(\ref{2}) imply that the mass of the dilaton $m_{\rm PGB}=
\left.\frac{\partial^ 2V }{\partial r^2}\right |_{r=\mu=\langle r\rangle, \langle\theta_i\rangle}$ 
arises at the  two-loop level,    
\begin{eqnarray}
m_{\rm PGB}^2 = 8C (\mu = \langle r \rangle) \langle r \rangle^2~.
\label{3}
\end{eqnarray}
Clearly, $C (\mu = \langle r \rangle) > 0$ is required.

The renormalization scale independence of the effective potential means that 
\begin{eqnarray}
\mu\frac{dV}{d\mu}\equiv \left( \mu {\partial \over \partial \mu} + 
\sum_{a} \beta_{a} {\partial \over \partial g_{a}} 
 +\sum_{x}\gamma_{x} m_x {\partial \over \partial m_{x}}- \gamma_r r {\partial \over \partial r}- 
 \sum_{i} \gamma_i \theta_i{\partial \over \partial \theta_i} \right) V = 0~,
\label{4}
\end{eqnarray}
where $\beta_a$ are beta-functions which determine the running of couplings $g_a$, 
while $\gamma_r$, $\gamma_i$ are scalar anomalous 
dimensions and $\gamma_{x}\equiv {\mu \over m_{x}}{\partial m_{x} \over \partial \mu}$ 
are mass anomalous dimensions.  Equations (\ref{1}), (\ref{2}) and (\ref{4}) imply
\begin{eqnarray}
B (\mu = \langle r \rangle) &=& 
\left. {1 \over 2} \mu {d A \over d\mu}\right |_{\mu = \langle r 
\rangle}, \nonumber \\ 
C (\mu = \langle r \rangle) &=& \left. {1 \over 4} \mu {d B \over
d \mu}\right |_{\mu = \langle r \rangle}~.
\label{5}
\end{eqnarray}
In principle, the $A$, $B$ and $C$ terms can be calculated in perturbation theory. The leading-order contributions to
$A$, $B$ and $C$ arise at tree $``(0)"$, one loop $``(1)"$, and two loops $``(2)"$, respectively.
If we are in the perturbative regime, then
the conditions $A(\mu = \langle r \rangle) = 0, \ B(\mu = \langle r \rangle) = 0$ and
$C(\mu = \langle r \rangle) > 0$ suggest:
\begin{eqnarray}
A^{(0)} (\mu = \langle r \rangle) \approx 0~, \nonumber \\
B^{(1)} (\mu = \langle r \rangle) \approx 0~, \nonumber \\
C^{(2)} (\mu = \langle r \rangle) > 0~.
\label{7}
\end{eqnarray}
The first condition can be used to estimate the scale $\mu = \langle r \rangle$,
resulting in the elimination of one of the tree-level parameters in the potential.
The quantity $B^{(1)}$ is in general
\begin{eqnarray}
B^{(1)} (\mu = \langle r \rangle) =\left. {1 \over 64 \pi^2 
\langle r \rangle^4} [3{\rm Tr} m_V^4 + {\rm Tr} m_S^4 - 4{\rm Tr}
m_F^4]\right |_{\mu = \langle r \rangle}~,
\label{8}
\end{eqnarray}
where the subscripts $V$, $S$ and $F$ refer to the contributions of massive vector bosons, 
scalars and Dirac fermions, respectively. It follows from Eq.(\ref{5}) that $C^{(2)}$ is given by
\begin{eqnarray}
C^{(2)} (\mu = \langle r \rangle) =\left. {1 \over 64 \pi^2 \langle r \rangle^4} \left[ 3 {\rm Tr} m_V^4 \gamma_V + {\rm Tr}
m_S^4\gamma_S - 4 {\rm Tr} m_F^4 \gamma_F \right]\right |_{\mu = \langle r \rangle}~, 
\label{9}
\end{eqnarray}
where $\gamma_x = \partial ln m_x/\partial ln \mu$ ($x = V, S, F$).

\emph{ A priori}, $C^{(2)}$ in Eq.(\ref{9}) need not be positive, 
thus the condition  $C^{(2)}>0$ potentially puts a restriction on the parameters of the theory. The condition $B^{(1)}\approx 0$ leads to the fermion-boson mass relation:
\begin{equation}
\left(3{\rm Tr} m_V^4 + {\rm Tr} m_S^4 - 4{\rm Tr}
m_F^4\right)\vert_{\mu = \langle r \rangle}\approx 0~.
\label{massrel}
\end{equation}
Higher loop corrections to this mass relation can be, of course, calculated within a given model. 
Thus, the above mass relation is an interesting
prediction of scale invariant models which arises by requiring the CC to be small (as suggested by observations).

The cosmological constant is a relevant observable only in the presence of gravity.
Gravity of course requires a mass scale, the Planck mass.
Let us briefly mention how such a mass scale can be generated within the scale invariant framework.
A simple way is to assume that the Planck mass arises through the couplings $\sqrt{-g}\xi_{ij}S_iS_j R$, where $R$ is the Ricci scalar.  
Because we assume that scale invariance is broken radiatively at the electroweak scale, to explain the weakness of the gravitational force one needs 
to introduce hierarchically large $\xi_{ij}$ parameters or, alternatively, 
one can invoke a large number of scalar fields. Another possibility, is that the Planck mass is 
spontaneously or dynamically generated in a hidden sector which extremely weakly interacts with the low-energy fields, 
rendering the effective low-energy theory to be approximately scale invariant. More detailed discussion  of these possibilities is beyond the scope of the present paper.

\section{The minimal scale invariant model revisited}

The simplest scale invariant model whereby electroweak symmetry breaking can occur is one which contains
a Higgs doublet ($\phi$) and one real singlet ($S$)\cite{Foot:2007as}.  
Let us apply the formalism outlined in the previous section to this model.
The most general scale invariant potential is:
\begin{eqnarray}
V_0 (S_1, S_2) = {\lambda_1 \over 2} \phi^{\dagger} \phi \phi^{\dagger} \phi + {\lambda_2 \over 8} S^4 + 
{\lambda_3 \over 2} \phi^{\dagger}\phi S^2 ~.
\end{eqnarray}
This potential might appear to have one more parameter than the standard model. In fact the
number of parameters is actually the same: both this potential and the standard model
potential have three parameters.  In the standard model case, there is 
the familiar quartic coupling $\lambda$ and the $\mu^2$ term; there is also a less
familiar `vacuum energy' parameter required to absorb the divergence in the vacuum energy
(see e.g.\cite{japan}). There is no such parameter in scale invariant models, since as we have just discussed, they have finite calculable CC. 

We parameterize the fields in unitary gauge through 
\begin{eqnarray}
\phi = {r \over \sqrt{2}} \left( \begin{array}{c}
0 \\
\sin \theta \end{array} \right), \ S = r\cos\theta~,
\end{eqnarray}
and we choose the $\lambda_3 < 0$ parameter space.
In this case, $V_0 (r) = A^{(0)} r^4$ and $A^{0} (\mu = \langle r \rangle) = 0, \ \partial A^{0}/\partial \theta = 0$ implies
\begin{eqnarray}
\sqrt{2}\langle \phi \rangle = \langle r \rangle \left( { 1 \over 1 + \epsilon}\right)^{1/2} \equiv v \simeq 246\ GeV,
\ \langle S \rangle = v\epsilon^{1/2},
\end{eqnarray}
where
\begin{eqnarray}
\langle \theta \rangle=\rho~,~~\cot^2 \rho  \equiv \epsilon = \sqrt{{\lambda_1 (\mu) 
\over \lambda_2
(\mu)}}~,
\label{aa6}
\end{eqnarray}
with
\begin{eqnarray}
\lambda_3 (\mu) + \sqrt{ \lambda_1 (\mu) \lambda_2 (\mu)} = 0
\end{eqnarray}
and $\mu = \langle r \rangle$.

The model has two physical scalars, but only one of them gains mass at tree-level.
The other scalar is the PGB, which gains mass at two loop level.
The tree-level mass can easily be obtained from the tree-level potential,
by defining shifted fields: $\phi = \langle \phi \rangle + \phi', \ S = \langle S \rangle + S'$.
We find:
\begin{eqnarray}
m_H^2 = \lambda_1 v^2 - \lambda_3 v^2, \ H = \cos \rho \phi'_0 - \sin\rho S',
\end{eqnarray}
while the PGB is $h = \sin\rho \phi'_0 + \cos\rho S'$.

As indicated in Eqs.(\ref{3}) and (\ref{9}), the mass of the PGB depends on the anomalous mass dimensions for the scalar $H$ and the top quark.
Evaluating these anomalous mass dimensions in the relevant parameter regime where $\cos^2 \rho \approx 1$, we find:
\begin{eqnarray}
\gamma_S &=& {3\lambda_1 \over 4\pi^2} 
- {9 \lambda^2_D \over 8\pi^2} \left( {m_{t}^2 \over m_{H}^2} - {1 \over 6}\right)
\nonumber \\
\gamma_F &=& {3\lambda_D^2 \over 32\pi^2} - {2 \alpha_s \over \pi}\ .
\end{eqnarray}
Using $\lambda_D^2 = 2m_{t}^2/v^2$ and the relation Eq.(\ref{massrel}), we find
\begin{eqnarray}
C^{(2)} =  {3m_t^4 \sin^4 \rho \over 16\pi^2 v^4} \left(  {2 \alpha_s \over \pi} + 
{3m_t^2 \over v^2 \pi^2} \left[ {3\sqrt{3} \over 8} + {1 \over 16}\right]\right)~.
\end{eqnarray}
Evidently $C^{(2)} > 0$ 
and hence the model is consistent with the inferred small CC.   The PGB mass can then be
estimated from Eq.(\ref{3}):
\begin{eqnarray}
m_{\rm PGB} &\approx & \sqrt{{3\over 2}} {m_t^2 \over \pi v}\sin \rho 
\left[ {m_t^2 \over \pi^2 v^2} \left( {9\sqrt{3} \over 8} + {3 \over 16}\right)
+ {2 \alpha_s \over \pi} \right]^{1/2} \nonumber \\
& \approx & 7 \left( {\sin \rho \over 0.3}\right) \ {\rm GeV}\ .
\end{eqnarray}
For such a light PGB, LEP bounds limit $\sin \rho \lesssim 0.3$ \cite{Abbiendi:2002qp}.

As discussed already, the incorporation of a small CC into scale invariant
theories implies some constraints on parameters. The main constraint is that $B (\mu = \langle r \rangle) = 0$.
To leading order in perturbation theory, this implies that $B^{(1)} \approx 0$ 
which leads to:
\begin{eqnarray}
m_H^4 \approx 12m_t^4 \ .
\label{bla2}
\end{eqnarray}
Note that the above relation is evaluated for the running masses at a scale $\mu = \langle r \rangle$. 
This relation leads to a predicted Higgs pole mass: 
$M_H\approx 280-305$ GeV for  $ 300\ {\rm GeV} \stackrel{<}{\sim} \langle r \rangle \stackrel{<}{\sim}  1$ TeV.
For large values of $\langle r \rangle \gg $ TeV the perturbative approximation potentially  breaks down due to the presence
of large logarithms in the perturbative expansion of the $A, B$ and $C$ terms in the effective potential, Eq.(\ref{a4}) . 

Higgs masses around 300 GeV are now disfavoured by LHC data\cite{lhc1}. Additionally the experiments have
found evidence for a Higgs-like particle with mass around 125 GeV\cite{lhc2}.
We conclude
that the simplest scale invariant electroweak symmetry breaking model appears to be
strongly disfavoured. 
The simplest model did not incorporate neutrino masses or dark matter.
These issues motivate extensions of the minimal model which we show can be fully realistic.

\section{Neutrino masses in scale-invariant models}

Incorporating neutrino masses within scale invariant models has been discussed in ref.\cite{Foot:2007ay} without regard to the CC constraints.
As discussed there,
the simplest way to generate Dirac neutrino masses is to introduce right-handed 
neutrinos which allow for tiny Yukawa couplings.  One could also consider a model with type I see-saw where the  right-handed 
Majorana neutrino masses are generated through the couplings with the singlet scalar field of the minimal model. Similarly, one 
could generate Majorana masses for  triplet fermions and generate light neutrino masses via the type III see-saw mechanism. Both type I and III models  
introduce new heavy fermionic states. 
However, with the Higgs boson mass fixed to 125 GeV, we need additional heavy bosonic degrees of freedom
to balance the left-hand side of Eq.(\ref{massrel}).
This motivates the type II see-saw mechanism as it introduces only new bosonic degrees of freedom. 
While type I and III see-saw models are fairly trivial modifications of the analyses of the 
previous section because they do not require any new scalars, type II is non-trivial in this respect.    
We now discuss the type II see-saw scale-invariant model in more detail. 

We extend the model of the previous section by introducing the triplet scalar field 
$\Delta  \sim (1, 3, -2)$ under $SU(3) \otimes SU(2)_L \otimes U(1)_Y$
\cite{Foot:2007ay}. 
If the neutral component of $\Delta$ gains a VEV then the neutrinos can acquire Majorana mass from
the Lagrangian term
\begin{eqnarray}
{\cal L} = \lambda \bar \ell_L \Delta \tilde{\ell_L} + H.c.,
\label{term4}
\end{eqnarray}
where $\tilde{\ell_L} \equiv i\tau_2 (\ell_L)^c$.
The simplest scale-invariant model which can give
$\langle\Delta^0\rangle \neq 0$ requires $\phi$, $\Delta$ and real gauge singlet scalar field, 
$S$ \cite{Foot:2007ay}.
The most general tree-level potential is
\begin{eqnarray}
V_0  &=& \lambda_1 (\phi^{\dagger}\phi)^2 + \lambda_2 ({\rm Tr}\Delta^{\dagger}\Delta)^2 +
\lambda'_2 {\rm Tr}(\Delta^{\dagger}\Delta\Delta^{\dagger}\Delta) + 
{\lambda_3 \over 4} S^4 + \lambda_4 \phi^{\dagger}\phi {\rm Tr}\Delta^{\dagger}\Delta 
+ \lambda'_4 \phi^{\dagger}\Delta \Delta^{\dagger} \phi \nonumber \\
& + & \lambda_5 \phi^{\dagger}\phi S^2
+ \lambda_6 \Delta^{\dagger}\Delta S^2 + \lambda_7 \phi^{\rm T}i\tau_2 \Delta \phi S + H.c.
 \end{eqnarray}
The only term which violates lepton number is the $\lambda_7$ term. This term will induce small
nonzero VEV for the neutral component of $\Delta$, provided that the
parameters are such that $\langle \phi_0 \rangle \neq 0$, $\langle S \rangle \neq  0$ and $\langle \Delta \rangle = 0$ when $\lambda_7 = 0$
\cite{Foot:2007ay}.
[An example of such a parameter choice is where all the $\lambda$'s are positive except for $\lambda_5$].
Minimising the tree-level potential in the limit $\lambda_7 \to 0$ leads to the relations,
\begin{eqnarray}
\lambda_5 (\mu) = - \sqrt{\lambda_1 (\mu) \lambda_3 (\mu)}
\end{eqnarray}
and 
\begin{eqnarray}
{\langle S \rangle^2 \over \langle \phi_0 \rangle^2 } = 
{w^2 \over v^2} = \sqrt{{\lambda_1 (\mu) \over \lambda_3 (\mu)}}.
\end{eqnarray}
A small but nonzero $\lambda_7$ induces order $\lambda_7^2$ corrections to these formulas.
As before we take $\mu = \langle r \rangle$.

We can calculate the tree-level masses by expanding around the vacuum: $\phi = \langle \phi \rangle + \phi'$,
$S = \langle S \rangle + S'$ and $\Delta = \langle \Delta \rangle + \Delta'$. 
Taking the limit $\lambda_7 \to 0$ we find that 
the physical scalar spectrum consists of an approximately degenerate complex
$\Delta'$ triplet, a massive Higgs-like scalar $H = \cos \rho \phi_0' - \sin \rho S'$, and a massless state 
$h = \sin \rho \phi_0' + \cos\rho S'$ (this is the PGB which will gain mass at two-loop level), where
\begin{eqnarray}
\cot^2 \rho &=& \sqrt{{\lambda_1 \over \lambda_3}}, \nonumber \\
m_{\Delta}^2 &=& {\lambda_4 \over 2} v^2 + \lambda_6 w^2, \nonumber \\
m_H^2 &=& 2\lambda_1 v^2 - 2\lambda_5 v^2.
\end{eqnarray}
Note that
the light PGB $h$ is not expected to contribute significantly to the width of the Higgs boson, $H$, despite
the fact that $H \to hh$ and $H \to hhh$ are kinematically allowed.
This is because the effective couplings
$Hh^2$ and $Hh^3$ vanish at both the tree-level and one loop level when evaluated at the renormalization scale $\mu = \langle r \rangle$. 
This is a general result for the couplings of the PGB.
To see this, recall from section 2 that the potential, Eq.(\ref{a4}), is constrained to have $A = B = 0$. This constraint forces 
the minimum to be on a flat direction (at one loop level). It also  
eliminates the $Hh^3$ and $Hh^2$ couplings (at one loop level) since otherwise movement along the flat direction, $h \to h + constant$ would
generate linear terms in $H$. Such terms would imply $\partial V/\partial H \neq 0$ and thus clearly cannot be present at the minimum.

As reviewed in section 2, the incorporation of a small CC constrains the effective potential terms
[defined in Eq.(\ref{a4})] to satisfy $C > 0$ and $B=0$ at the renormalization scale $\mu = \langle r \rangle$. 
As in the minimal model of the previous section, one can easily check that the $C^{(2)} > 0$
does not impose any significant constraints on the parameters of the model due to the large positive top
quark contribution to $C^{(2)}$.
The constraint $B=0$ leads to the mass relation (valid at $\mu = \langle r \rangle$):
\begin{eqnarray}
6m_\Delta^4 + m_H^4 \approx 12 m_t^4 \ .
\end{eqnarray}
The $\Delta$ states are not expected to contribute to oblique radiative corrections, and, therefore, they can be heavier than $m_H$. 
Assuming $m_H \simeq 125$ GeV, then we end up with a prediction
for $m_\Delta$ of :
\begin{eqnarray}
m_\Delta \approx 200\ {\rm GeV}.
\end{eqnarray}
Importantly, this puts the mass of the $\Delta$ scalars in the range where they can be probed at the LHC.

The PGB is light $\sim$ GeV and can be produced in association with top quark pairs or $W,Z$ pairs
at the LHC:
$pp \to \bar t t h, \ pp \to ZZ h, \ pp \to WW h, \ pp\to WZh$.
A particularly striking signature will occur if $m_h < 2 m_\tau$. In this case $h$ will have 
significant branching fraction to muon pairs, $h \to \bar \mu \mu$ (or $h \to \bar e e$ if $m_h < 2 m_\mu$).
Also, in this parameter range $h$ can potentially live long enough to produce a displaced vertex. Detailed
phenomenological studies of this kind of PGB are clearly warranted which we leave to future work.
 
\section{Hidden sector dark matter}

A variety of observations require non-baryonic dark matter in the Universe\cite{review}.
Hidden sector dark matter is a theoretically attractive option since
hidden sector particles are automatically dark and their stability can be readily achieved
via `accidental' global symmetries.
Importantly such dark matter is virtually unconstrained
by current collider searches. Furthermore these particles can be easily
light and able to explain the DAMA\cite{dama}, CoGeNT\cite{cogent} and CRESST-II\cite{cresst} data.
See e.g. ref.\cite{hs} for recent discussions of hidden sector models in this context.

Electroweak scale invariant models require at least one electroweak singlet Higgs, $S$.
If dark matter resides in a hidden sector, then electroweak singlet Higgs states are required to 
give mass to the dark matter particles.  Bare mass terms are forbidden by the scale invariance symmetry. 
It is quite natural therefore, to associate $S$ with
the hidden sector. This will also have the virtue of anchoring the dark matter mass scale to the electroweak scale,
which is especially important in light of the results from the direct detection experiments.

We illustrate hidden sector dark matter within the scale invariant setting by considering the
simplest non-trivial Hidden sector with $U(1)'$ gauge symmetry.    
If $S$ is charged under the $U(1)'$ gauge symmetry, then
$\langle S \rangle$ will break this hidden sector gauge symmetry. This VEV can also
be responsible for the masses of the dark matter particles if
the scalar $S$ couples to hidden sector fermions. The simplest fermionic content is comprised of a chiral pair of fermions, $(F_{1L},~F_{1R})+(F_{2L},~F_{2R})$, that transform as vector-like representation of $U(1)'$ gauge group: $(q, q-1)+(q-1,q)$, where we have 
normalized   $U(1)'$ charge of $S$ to unity without loss of generality. Being the vector-like representation, 
the fermionic content is free of $U(1)'$ as well as mixed $U(1)'$-gravity anomalies. For $q\neq 0$ and $q \neq 1$ the following Yukawa interactions are allowed: 
\begin{eqnarray}
{\cal L} =  \lambda_1 S \bar F_{1L} F_{1R} \ + \lambda_2 S^{*} \bar F_{2L} F_{2R}  + {\rm H.c.}~, 
\end{eqnarray}
which generate Dirac masses for $F_1$ and $F_2$ upon $U(1)'$ gauge symmetry breaking. Interestingly, the Lagrangian
exhibits an `accidental' $U(1)_1\times U(1)_2$ exact global symmetry, which guarantees that both hidden sector Dirac fermions are stable
and can therefore constitute dark matter\footnote{If $q=0$ or $q = 1$, additional Yukawa couplings are allowed that result in 
the hidden sector fermions gaining Majorana masses. In this case the global $U(1)_1\times U(1)_2$  symmetry is broken. Nevertheless, there remains an
accidental $Z_2$ symmetry, generated by hidden sector matter parity, $H: F_{iL(R)}\to -F_{iL(R)}$.
This  symmetry ensures that the lightest hidden sector fermion is stable and thus could account for dark matter.}. 

Recall that the incorporation of a small CC constrains the effective potential terms
[defined in Eq.(\ref{a4})] to satisfy $C > 0$ and $B=0$ at the renormalization scale $\mu = \langle r \rangle$. 
As before the condition $C^{(2)} > 0$ is easy to satisfy. 
The constraint $B=0$ leads to the mass relation (valid at $\mu = \langle r \rangle$):
\begin{eqnarray}
m_{Z'}^4 \approx 4 m_t^4
\end{eqnarray}
where we have assumed that the hidden sector fermions are much lighter than the top quark.
This relation suggests the mass of the $U(1)'$ gauge boson $Z'$ is $m_{Z'}\approx 240$ GeV in this 
minimal hidden sector model.

The discussion above serves to illustrate our main points, which are (a) that hidden sector dark matter can be
very naturally incorporated within scale invariant models and (b) such models can fix the dark matter
scale to the weak scale with specific predictions arising from the mass relation, Eq.(\ref{massrel}).
It is, of course, straightforward to extend this analysis to more complicated hidden sectors. 
In particular, scale invariant models with
larger hidden sector gauge symmetry which feature additional light or massless $Z'$'s can be straightforwardly
constructed. These include models realizing the specific scenarios discussed in the recent literature\cite{hs}.

\section{Conclusion}

Scale invariance is a promising candidate for a symmetry of nature at some level.
It is certainly interesting to wonder whether the effective theory
at energies below the Planck scale might be (classically) scale invariant
broken perturbatively only by quantum corrections.
This is the idea envisaged by Coleman and Weinberg many years ago\cite{Coleman:1973jx}.

Such scale invariant theories have a number of interesting features. Firstly, scale invariance
can protect the weak scale from radiative corrections. Scale invariance is broken by the anomaly
but this breaking is ``soft'' as it generates only logarithmic corrections to the weak scale\cite{bardeen,Coleman:1973jx}.
Thus scale invariant models of the kind considered here provide an interesting solution to
the so-called hierarchy problem. Another interesting aspect of scale invariant theories is that
the cosmological constant is a finite calculable parameter. Setting it to be small, as suggested
by observations, leads to interesting
constraints on the parameters\cite{Foot:2010et}. 
This is quite unlike the standard model where it is divergent requiring an aribitrary counter term
and thus does not lead to any parameter constraints.
 
We have examined the simplest scale invariant models with small cosmological constant.
These models are highly predictive.
Indeed, the  minimal such model, comprising a Higgs doublet and a real singlet, has the same
number of parameters as the standard model.  It turned out, though, that this minimal model is disfavoured by recent
LHC data. Of course the minimal model, like the standard model, requires modifications to 
explain non-zero neutrino mass and dark matter.
We therefore examined simple extensions to the minimal model 
which incorporate this new physics.  
We found that scale invariant versions of the type-II see-saw model and hidden sector
dark matter are fully realistic. That is, consistent with all experiments and observations.
These models predict a light pseudo-Goldstone boson, $h$,
with mass around 10 GeV or less. A fermionic-bosonic mass relation is also predicted.
These specific models, as well as more generic scale invariant models, can be probed at the LHC.

\subsection*{Acknowledgements}

We would like to thank Ray Volkas for collaboration during the early stage of this work.   
The work was supported in part by the Australian Research Council.

\end{document}